# Big Data and Social/Medical Sciences: State of the Art and Future Trends

Adil E. Rajput[1] and Samara M. Ahmed[2]

[1]College of Engineering, Effat University, Jeddah.
[2] College of Medicine, King Abdulaziz University, Jeddah.

**Abstract.** The explosion of data on the internet is a direct corollary of the social media platform. With petabytes of data being generated by end users, the researchers have access to unprecedented amount of data (Big Data). Such data provides an insight into user mental state and hence can be utilized to produce clinical evidence. This lofty goal requires a thorough understanding of not only the mental health issues but also the technology trends underlying the Big Data and how they can be leveraged effectively. The paper looks at various such concepts, provides an overview and enumerates the work that has been done in this realm. Furthermore, we provide guidelines for future work that will help in streamlining the Big Data use in social/medical sciences.

## 1 Introduction

The social media platform presents a set of challenges to practitioners in the medical and social sciences. The opportunity to meet people from various backgrounds, ages and location allows users to exchange information in various contexts. This exposes the users to different stimuli that can have adverse effects on mental health of certain individuals (especially if they are susceptible to certain criteria)**.** The recent Parkland shooting in US shows the alleged shooter was exposed to and influenced by a hate group on the Internet. In addition, the shooter had depicted many signs and symptoms of mental health issues including anger management that went unnoticed. Practitioners in such instances find it difficult to gather specific data from huge number of posts.

However, the aforementioned challenge is accompanied by unprecedented opportunities. The advent of Big Data techniques has provided practitioners in various fields to sift through multitude of data to infer insights that were once unavailable. Such insights range from general observations regarding group of people (e.g., teenagers' trending interests) to specific behavior concerning a particular individual or event (e.g., use of sarcasm in a particular tweet). The Big Data science is characterized by the three characteristics namely Velocity, Veracity and Variety [14].

Both Psychologists and Psychiatrists have started to reckon and realize the benefits of this trend. The past five to seven years have seen increasing application of Big Data techniques to social and medical sciences. One of the challenges faced by the practitioners is for both sides to gain a better understanding of the other domain. In other words, Computer Scientists and Psychologists/Psychiatrists have started developing expertise beyond their respective domains. The literature shows quite a good amount of work that has conflated the ideas from both these fields. However, the work is sporadic when plotted against the scientific concepts that underlie the Big Data field. One reason for this is the fact that Big data field is a confluence of many traditional fields in Computer Science namely Artificial Intelligence, Machine Learning, Data Mining, Natural Language Processing and Information Retrieval. Furthermore, the application of Big Data concepts to medical and social sciences lack a systematic approach. While the work in the field of Bioinformatics has been thorough, the same cannot be said about the mental health domain. The work ranges from detecting sarcasm to sentiment analysis.

The work done pre-social media is based on standard corpus, which required the usage of standard vocabulary words found in a dictionary. For English language the literature included both British and American English. However, the social media platform specifically and Internet in general added the Out of Vocabulary (OOV) words which did not conform to a particular linguistic style and present in any dictionary. The global usage of social media also complicates the matter further as a particular platform will have not only different styles of English language usage (same probably holds true for other languages) but also colloquial words specific to different ethnicities e.g., Spanglish, Hinglish etc.

Looking at the American Psychiatric Association guidelines published in DSM V, one of the goals is to make Psychiatry more of an evidence-based field. The social media platform provides a plethora of data that researcher can access and study from various angles. The work in this paper aims to give a general overview of the status quo and the ripe opportunities.

**1.1 Objectives**

The paper will present the following
1. Overview of various fields underlying the Big Data application to Psychology/Psychiatry
2. State of the art
3. Areas where further work can be accomplished as a natural corollary of existing work

**2 Natural Langue Processing (NLP)/Information Retrieval (IR)**

NLP is a major area in the field of Machine learning where the goal is to process natural human language and infer meaningful insights. Specifically, Machine Learning refers to the ability of machines to acquire intelligence over a period of time that comes naturally to humans. The ELIZA system developed at MIT in the mid-1960s [3] - interestingly built to mimic a Rogerian psychotherapist (focused on helping client's self-actualization tendency). As an example, the system asked a human who complained of a headache to consult a doctor. Similar ideas are still in play with various chatbots entering the scene and lately with the popularity of Siri and Alexa. The underlying technique remains that of pattern matching where the term "headache" was annotated with consulting a doctor. The patterns specifically refer to text strings that are usually annotated. The problem becomes interesting when patterns consist of more than one word. Thus a regular expression such as "long bouts of insomnia" needs to be searched for as a whole and not as separate word when it comes to mental illness.

The traditional approach to NLP takes a systematic approach of syntax, semantics and pragmatics. The syntactical/lexical analysis simply parses the semantics into words (tokens) while the semantic analysis puts the words in context. This is where words such as "depress", "depression" and "depressed" are pooled together to as synonyms. Stemming is such an approach introduced in [4] and refined over the years. The pragmatic part provides annotation to certain words in a context. For example, "feeling blue" is considered synonymous to being depressed. Such techniques depend on having a sizeable literature out of which keywords are extracted. One of the popular methods remain the term frequency model (tf-idf) described in [2].

Statistical approach became popular with NLP where a sample representation of text was considered a sufficient sample to represent the whole literature - the basic premise being that the keywords extracted from a large enough sample will not increase significantly in size compared to the entire archive. The work done in this domain by linguistics is described in [23]. One of the advantages of this approach was that not all the documents were available to researchers on line and searching improved gradually as the size of corpus increased.

The techniques provided in the NLP field provided enough to help practitioners in the social/medical fields to conduct their work without understanding the technical details. Traditionally, the libraries provided a way to search specialized documents. The advent of World Wide Web propelled the field to new heights when researchers started working in the area of Information Retrieval (IR). Even though the scope of IR encompassed all types of information, the work was primarily focused on handling text queries. The main premise is that both the user query and the sample representation of the literature will be incomplete. Similar to NLP, the IR builds a list of keywords - as indexes - to search. This is done by removing stop words such as "and", "or" etc. and performing a normalization process using techniques such as stemming. This also helped in developing ontologies and recognizing phrases such as "feeling depressed" and equating it to the underlying symptoms. Search engines such as Google did an excellent job of retrieving relevant documents given a search query including terms used by psychology practitioners and hence never required social sciences and medical practitioners delve in the fields described here. The two measures that IR uses are Precision and Recall. Precision measures the fraction of retrieved items that were relevant whereas Recall measures fraction of relevant items retrieved.

The DSM V [27] guidelines focus in transitioning to be a more evidence based field. One of the challenges faced was the lack of data for myriad of reasons including the stigma tied to mental illness. The social media offers the researchers many ways to dissect the data available to them. This can include separating the posts based on location thus narrowing it down approximately to a certain ethnic type. The next section will discuss this in more detail.

## 3 Big Data and Social Media

Both Psychologists and Psychiatrists are concerned with pinpointing various factors during the patient interview phase that can help them better diagnose and treat the underlying problem. The issue is compounded as many mental illness health conditions come with comorbidity. Techniques described in [10] help pinpoint the user location. Such techniques have been used for different reasons that include discovering upcoming social events [12] and providing emergency services [9]. Moreover, the work described in [40] helps identify the gender of the user making the post. These techniques can be very beneficial to researchers when discussing behaviour of certain cohorts based on certain criteria.

One of the problems faced above is the "noisy" language that is used in the social media [25]. In other words, the social media users do not conform to the standard words that are present in a dictionary. Hence, it would be difficult to recognize "dprssn" as depression. Furthermore, teenagers and adolescent appear to not conform to a particular style of vocabulary. Rather, the patterns change over a period and introduce new Out of Vocabulary words.

The challenges posed to researchers working on detecting mental illness or certain symptoms is twofold:
1. Can social media platform provide a scientific sample that can be used to gain insight into users' behavior
2. How to deal with Out of Vocabulary words

We discussed the importance of having a corpus in section 2. There has been a significant work done in by Computer Scientists and Linguists - both historically and recently - that discussed various properties of corpus. These include:

1. Providing a context improves the precision and recall of the queries [22]
2. Weblogs can be used to narrow down the subject at hand [26]
3. The representative corpus can be one-fifth of the reference corpus [23]

So far there has been not any focus on providing a sample corpus specific in psychology/psychiatry. Authors in [1] provide a first step in establishing the need of such a corpus and provide initial experimental results confirming results from the literature. However, researchers have many avenues that can be explored to 1) establish and update a corpus that can be used for social media usage scientifically and 2) devise a mechanism to associate OOV words that can potentially indicate presence of an underlying symptom with certain accuracy. One approach could be to gather a sample corpus by people with certain mental illness and comparing it against a standard corpus from DSM V. Such a step can be imperative in furthering work in this field.

## 4 Applications in Psychology/Psychiatry

There has been quite some work that has been done in this realm. Work done by [24] provides a framework that helps in extracting Big Data from the Web. The authors specifically promote a "theory-driven" web scraping where researchers are encouraged to formulate questions and hypotheses before embarking on scavenging data from the web. This can include various criteria such as age, demographics etc. One of the criticism of this work is that while this can be great for supervised learning (where a training dataset is present), it might not be that effective in unsupervised learning scenarios – where we need to sift through multitude of data to classify and "cluster" similar information together. The authors also present a case study where they utilize the gender identity to prove a hypothesis regarding women behaviour. Lastly, the authors point out the importance of ensuring about the legality of using the data. The work provides researchers an excellent starting point about how web crawlers work and what role an Application programming Interface (API) plays in Big Data. Another important aspect to keep in mind is storing of the data. Traditionally data has been stored in relational databases such as Oracle, SQL Server etc. However, NoSQL databases are better suited for such applications. MongoDB is a popular choice where researchers can view/analyse data from various dimensions.

Work done by [18] focuses on the Facebook platform where the authors compare the human perception to one being gleaned from the social media. Specifically the authors measure users' judgements by the numbers of "likes" the users press. They build two samples of more than 14 thousand users where they ask the users' friends to rate the user judgement. The results showed that Facebook likes provided a more accurate view confirming the hypotheses that information obtained automatically from social media is both valuable and accurate. The results do not reflect the comments that the users make on various sites/items which can grant more accuracy to the results. Moreover, there is no corpus involved in the process.

Work done by [15] provides an overview of Big Data application to Psychology. The authors focus on the four steps necessary for such endeavours namely planning, acquisition, planning and analytics. They also provide three tutorials for the users. They introduce the user to the MapReduce framework that has garnered lot of attention lately. The paper also explains the concept of supervised and unsupervised learning as explained above. The work while providing an excellent overview does skip certain details that the user might need. Specifically, they glance over the pre-processing part of data processing and the many underlying details such as text normalization.

The work done by [13] focuses on text normalization for the Out of Vocabulary words. Specifically, the users target words such as "smokin" and find a mechanism to convert it into "smoking". The work focuses primarily on SMS messages and also delves into a decent sample size for Twitter. The authors' proposed method produces very encouraging results comparing it to a corpus obtained from New York Times. The OOV words are first normalized using a dictionary-based matching and based on the results the authors move further to test in a context setting.

The authors work is mimicked in [1] where the work shows results in a psychology/psychiatry context – specifically depression. One of the issues the authors do not look into is converting OOV synonyms into actual dictionary based words – as an example "imo" into "in my humble opinion". A similar approach of normalization was also taken by [26] where the authors construct a normalization dictionary for Weblogs. The work provides an excellent way of pre-processing blogs – another social media platform. The work is not domain specific and hence the ideas can be applied to a domain specific context.

Work done at Stanford University [29] looks at the concept of digital footprint of a user and two mathematical ways to analyse data. The work uses R language (as opposed to Python) and helps predict real life outcomes. The work is based in the unsupervised learning domain and uses Facebook data as a case study. While the work is not focused in the medical/social sciences domain, it is an excellent introduction to digital footprint of a user and can come in very handy in detection of symptoms specific to various conditions.

De Choudhry et. al. has done some work in detecting mental illness. The authors started the work by focusing on detecting post-partum depression. The authors chose reddit as their platform and studied the linguistic changes that happened in new mothers. They showed the prevalence of negative affect in certain cases among other results. The work was an excellent first step in this realm and was followed by [20] where they predict depression in Twitter users. One of the prerequisite of their work is that the users identify themselves as depressed and gave consent to follow their Twitter account. The next step in this field should be to detect symptom of depression from random set of tweets. Moreover, the DSM V guidelines should also be followed and brought in line with the social media text.

De Choudhry et. al also did some work [8] in focusing on the anonymity factor of the social media and the disinhibitions that accompany such anonymity. Specifically, they looked at the 'throwaway' accounts that the users used to express their opinions. The authors also applied more Big Data and Machine Learning techniques to the Nutrition area where they used social media to understand dietary choices among the social media users [7]. While the work is not directly related to mental illness, it can be leveraged in various situations such as cases of eating disorders occurring with other conditions such as depression or bipolar.

Saha et. al. modelled stress among a group of college students after cases of shooting on campus [6]. The authors used reddit campus community and looked at the linguistic style of the students making posts to detect high level of stress after such traumatic incidents. They looked at 12 incidents of gun violence over a period of five years and analysed both the time and linguistic dimension of the posts. The work done gives a great platform to spring from and see whether clinical inferences can be done given such data.

It is worth mentioning the work in [5] where the authors choose YouTube platform and focus on detecting spam in the comments section. The authors use graph and network theory to look at bot (automatic programs) behaviour. Such work can be combined with other work and help researchers classify various videos from myriad of dimensions. Once such classification is accomplished, the researchers can look at the embedded comments and tie the content to user mental state and how they are expressing it.

While the aforementioned work analysed from a generic point of view (applicable to both psychology and clinical part of psychiatry), [16] present a list of projects that are underway in the realm of psychiatry. The authors look at the Big Data field from a medical sciences perspective and emphasize how Big Data can provide benefits in a clinical setting such as pinpointing rare events. The work done in this area ranges from the use of psychotropic drugs to comparing the risk of dementia in certain age group. The work presented does not include many pertinent details that bridge the gap between Computer Science and psychiatrists.

## 5 Conclusion and Future Work

The work in this paper aims at providing an overview of Big Data, its underlying technologies and how they can help practitioners in medical/social sciences field to analyse multitude of data from various perspectives/dimensions. Such data can be very objective and provides a much larger sample space than what is usually available in clinical settings today. The work in this paper provides an overview from both technological and clinical perspectives.

Future endeavours include the following:
1. Perform a PRISMA review for mental health issues
2. Tying the current work to cultural psychiatry (looking at data specific to certain cohorts with a common set of lifestyle)
3. Apply certain techniques to areas of cyber bullying and suicide ideation